\RequirePackage[2020-02-02]{latexrelease}
\documentclass[prl,showpacs,preprintnumbers,amsmath,amssymb]{revtex4}

\usepackage[dvips]{graphics}
\usepackage{tikz}

\begin{document}

\title{Locality in the Schr\"odinger Picture of Quantum Mechanics}

\author{Vlatko Vedral}
\affiliation{Clarendon Laboratory, University of Oxford, Parks Road, Oxford OX1 3PU, United Kingdom}

\date{\today}

\begin{abstract}
\noindent We explain how the so-called Einstein locality is to be understood in the Schr\"odinger picture of quantum mechanics. This notion is perfectly compatible with the Bell non-locality exhibited by entangled states. Contrary to some beliefs that quantum mechanics is incomplete, it is, in fact, its overcompleteness as exemplified by different pictures of quantum physics, that points to the same underlying reality. 
\end{abstract}

\pacs{03.67.Mn, 03.65.Ud}

\maketitle                           

\section{Introduction}
The concept of the field was invented in physics in order to preserve the idea of locality. Fields came as an answer to Newton's (and many other people's) worries that gravity seems to act at a distance, since massive objects attract one another across the vastness of space without any (apparent) mediation in-between. In order to remedy this, Faraday came up with the idea of the field, so that an object only affects the field in its immediate vicinity, and this disturbance then propagates through the field much as a wave would when a stone is thrown into a pond. This wave of disturbance in the field ultimately reaches another object, and the field disturbance then affects it locally, at the point where the other object is (as a lotus leaf in the pond would be, after the wave caused by the stone reaches it). That way, any action at a distance is avoided. 

Exactly the same is true for quantum fields \cite{Schweber,Weinberg}, the only difference being that what does the waving in the quantum case are the q-numbers pertaining to the field \cite{Vedral-local,Vedral-mach,Vedral-double}. In the case of the electromagnetic field, these would be the operators characterizing the vector potential, the electric and the magnetic fields \cite{Kallen} (with subtleties involving gauges \cite{Scharf} which will not be relevant for the present work). Here the speed of propagation is of course the speed of light. But the speed itself is irrelevant for the concept of locality. Instead, we simply need to require that acting on one subsystem (whatever this is) cannot instantaneously affect another subsystem. We will spell out below exactly what this means, but, suffice it to say, this principle - known as Einstein's locality - is also true in non-relativistic quantum mechanics. If we have two qubits located at two different places, then acting on one of the qubits cannot change any property of the other qubit. 

It is sometimes said that quantum physics violates Einstein's locality. But as explained above, it doesn't. Quantum physics violates, an altogether different, idea of locality called Bell locality. In Bell locality, things are described by c-numbers (not q-numbers) locally. And, of course, we know that the violation of Bell's inequalities is in direct conflict with the existence of the c-number based local reality. Some confusingly conclude that the violation of Bell's inequalities means that quantum physics is Einstein non-local (instead of just Bell non-local), but this is only true if one assumes that reality is described by the c-numbers locally. However, given that quantum physics does satisfy Einstein's locality, it is more appropriate to say that quantum physics violates the c-number based reality (this too is unfortunately frequently overstated by saying that ``quantum physics proves that reality doesn't exist"). It simply means that reality is underpinned by locally-given q-numbers and this is a radically different form of reality from the classical one, but it is a reality nevertheless. Nothing in quantum physics says that there is no reality whatsoever. That statement would not make any sense even to a philosopher. 

It is important to discriminate between different notions of locality since sometimes we hear incorrect conclusions that quantum physics is incompatible with General Relativity because the former in non-local while the latter is local. This however confuses the two notions of locality. The two theories in fact agree perfectly, since both are Einstein local. The fact that quantum physics is Bell non-local and General Relativity is not, does not represent an obstacle to their unification (no more than it is an obstacle to quantum electrodynamics). When the fundamental entities in General Relativity (the metric tensor and the Christoffel symbols) are quantized, the resulting theory of Quantum Gravity also becomes Bell non-local. Any classical theory that is quantized is automatically - by its very construction - Bell non-local.  

Here we intend to target the following related issue (and show that it is not an issue). If quantum physics is local and q-number realistic, then why does the wavefunction (or, more generally, the density matrix) give us the impression of non-locality \cite{Deutsch}? Namely, if the state of two qubits is maximally entangled, say $|00\rangle + |11\rangle$, then introducing a phase shift on, say, the first qubit ($|1\rangle \rightarrow e^{i\phi}|1\rangle$, leads to the state $|00\rangle + e^{i\phi}|11\rangle$. But the latter state is indistinguishable from the state where the phase shift is introduced in the second qubit. It looks as though the phase ``travelled" instantaneously from the first qubit, where it was generated, to the second qubit, which could be arbitrarily far away. Note of course that there is no way to extract the phase information from the second qubit alone (or the first one for that matter - the key feature for the protocols known as quantum data hiding). In that sense there is no possibility of using this local phase kick to signal to the second qubit. Formally, the reduced density matrices of either of the qubit are independent of the phase (they are simply in a maximally mixed state each). But there is more to this. No property of the second qubit, i.e., no q-number locally pertaining to it, is affected by the phase kick in the first qubit.  

Indeed, when we are lured into thinking that something non-local may be going on, we are being deceived by the view that only the wave-function is relevant to the factual situation of the two systems. The reason is that the state of the system only tells us half of the story in quantum physics. We need the observables (including the state) as well as the dynamics of the system. Locality is all about the dynamics, and this is true in quantum physics as much as in the water wave propagation in the pond. There is also a way of expressing the state in the Schr\"odinger picture that retains the local knowledge of the phase by keeping track of all the operations executed on the system. But, of course, any given state does not have enough ``capacity" on its own to contain all of the dynamics. This paper is about this and related issues. 

We will now proceed to review the two main pictures of quantum mechanics, the Heisenberg and the Schr\"odinger, in order to show how locality is understood in each of them (it's more or less the same since the dynamics is the same in both). The rest of the article then analyses the basic quantum interference as well as entangled states from the local quantum perspective.

\section{The Heisenberg and Schr\"odinger Pictures}

Most of us are used to calculating in the Schr\"odinger picture simply because it is frequently easier to do so. However, the Heisenberg picture is much closer to our intuition which is based on learning Newton's classical mechanics in high school. The Heisenberg picture arises naturally in quantum field theory because we arrive at it by (second) quantizing the wave-function, which therefore becomes an operator in quantum field theory. In this manner, when the Schr\"odinger picture is used in the first quantization, it becomes automatically the Heisenberg picture upon performing the second quantization. The two pictures are nevertheless equivalent in much the same way as the Hamilton approach to classical mechanics is equivalent to Newton's. They simply lead to exactly the same experimental predictions. But even more, they both tell us the same thing about what ``really" is going on. 

More formally, in the Schr\"odinger picture, quantum states evolve in time, while observables are stationary. In the Heisenberg picture it is the other way round. It is because the observables evolve in the Heisenberg picture that this is closer to the spirit of classical physics, where the position or momentum of a particle change in time. It is, however, irrelevant what evolves and what doesn't (this is not real, one can't tell if the states or the observables evolve). The only empirically accessible quantity is of the form $\langle \psi |O|\psi\rangle$ and this quantity changes in time by either evolving the states $|\psi (t_0)\rangle \rightarrow U(t,t_0) |\psi (t_0)\rangle$ (Schr\"odinger) or the observables $U^{\dagger}(t,t_0) O(t_0)U(t,t_0)$. The expected value is the same since $(\langle \psi |U^{\dagger})O(U|\psi\rangle) = \langle \psi |(U^{\dagger}OU)|\psi\rangle$.

Geometrically, this can be seen as follows. States and operators are vectors in quantum mechanics. The physical meaning lies in their inner product, which in Euclidean geometry would be (cosine of) the angle between the vectors. Now, we can evolve one of the vectors in, say, the clockwise direction in the two-dimensional plane defined by the vectors, while keeping the other vector fixed. This will lead to an increased angle between the two and therefore a changed inner product. But, the same could be achieved by evolving the other vector by the same amount, but in the counter-clockwise direction, so that the angle between the two ends up the same as in the former case. This is precisely why, in quantum mechanics, if the density matrix evolves as $\rho \rightarrow U\rho U^{\dagger}$ (Schr\"odinger), the operator must evolve (``backwards in time") according to $O\rightarrow  U^{\dagger}\rho U$ (Heisenberg). 

We now give a related account of thinking about the Heisenberg picture that will help us understand the main argument. The idea will be that as the state is rotated, we also rotate the basis, so that the state ultimately stands still with respect to the rotated basis. This is the ``active way" of looking at the Heisenberg picture: the state evolves as in the Schr\"odinger picture but then this evolution gets compensated by the evolution of the basis so that the state is stationary (always with the same coordinates - amplitudes in quantum physics - with respect to the current reference frame). For any basis vector $|a\rangle$ let's define $|a_t\rangle=U^{\dagger}(t,t_0)|a\rangle$. Now, the time evolution of this state is $|a_t (t)\rangle =  U(t,t_0)|U^{\dagger}(t,t_0)|a\rangle = |a\rangle$. If we look at the spectral decomposition of an operator, $A = \sum_n a_n |a_n\rangle\langle a_n|$, then the evolution of the basis leads to $A(t) = \sum_n a_n |a_{n,t}\rangle\langle a_{n,t}|$ (the observables, if the coordinate system moves with the state, must appear to be moving backwards). When the Hamiltonian is time-independent, we have $U(t,t_0) = \exp\{-iH(t-t_0)\}$.  

Incidentally, all other infinitely many pictures in-between Schr\"odinger and Heisenberg are also equally good. We can evolve the states for half the time and the operators for the other half (backwards) and the resulting expected value (the vector inner product) would still be the same. The so called interaction picture is like that and so is, speaking somewhat loosely, the two state formalism of quantum mechanics. However, here we focus on just the Schr\"odinger and Heisenberg pictures as they are antipodal to one another and it would superficially seem that they are as different as it could possibly be. 

Let's return to the example of kicking in the phase to one qubit, which is part of a of two-qubit maximally entangled state. It looks as though the Schr\"odinger picture misleads us into believing that the maximally entangled state is non-local, namely that the two qubits are behaving non-locally in some way, because the phase looks like it could also pertain to the other qubit. In the Heisenberg picture instead, no such thing happens. Namely, if we think of the $X$ q-numbers of the qubits as $X_1 = X\otimes I$ and $X_2 = I\otimes X$, where $X$ is the usual Pauli $X$ matrix, then introducing the phase (suppose for simplicity that $\phi = \pi$) in the first qubit leads to $X_1\rightarrow -X_1$ while $X_2$ stays the same (a similar conclusion is reached for $Y$, while neither of the $Z$s change). The phase kick in the first qubit therefore clearly only affects the observables pertaining to the first qubit ($X$ and $Y$) and not the second one. 

Furthermore, the state in the Heisenberg picture can always be taken to be a product state between the two qubits and, because it never evolves in this picture, there are never any issues with entanglement. The state is also fully local. This last point is not directly relevant as the locality we are discussing, of the Einstein kind, is all about the dynamics. As we will see below, any entangled initial state can anyway be thought of as starting as a product state and the operations needed to create entanglement can then be absorbed into the evolution of the operators. The state is irrelevant in the Heisenberg picture because it does not change. One could choose it to be entangled and it would not affect the Einstein locality, because this locality is about what happens to other system when one dynamically changes the first system.  Altogether, this is what we mean by Einstein's locality, both states and operators reflect local operations locally and this is true in all pictures of quantum physics. 

So why does the Heisenberg picture give us the impression that it conforms to Einstein's locality fully while the Schr\"odinger's doesn't? Because, when working in the Schr\"odinger picture, we frequently (and mistakenly) forget to take into account the dynamics itself, when we discuss the evolution of states. We only present the final state of the system and ``forget about" the dynamics through which we got to that state (i.e. we write the phase down, but we do not keep the information regarding how it was induced). This is sufficient as a starting point for all the future calculations, but it does not contain the answer to the question of how things happened in the past (like where the phase was inserted). When the dynamics is fully included, not only do the two pictures give the same experimental predictions (since they lead to the same expected values of all quantities), but they also tell one and the same story, in a perfectly local manner. We turn to this topic next. 

\section{Entanglement and Locality in the Schr\"odinger picture } 

Here we will present the Schr\"odinger picture in a slightly different way that will be easier to apply to the problem at hand. As a warm up, let us first think of a familiar free evolution of a single particle. For a free particles all the operators can be taken to be functions of its position and momentum, $O(p,q)$. This is true for the density matrix too $\rho (p,q)$. 

In the Heisenberg picture we can assume that $\rho (p_0,q_0)$, where $p_0,q_0$ are the initial momentum and position of the particle. The dynamics, just as in classical mechanics is given by $p(t)=p_0$ and $q(t) = q_0 + p(t)/m t = q_0 + p_0/m t$. The momentum operator never changes (since the particle is free), while the position operator becomes a function of the initial position plus $vt$, just like in classical physics. Here the observables change in time and acquire (in general) an explicit time-dependence through $O(p(t),q(t))=O(p_0,q_0 + p_0/m t)$.

To get to the Schr\"odinger picture, we assume that observables are stationary while the states now acquire an explicit time dependence. In that sense, $O(p_0,q_0)$ while $\rho (p_0,q(t)-p(t)/m t)$. Note the minus sign in the evolution of the density matrix in the Schr\"odinger picture with respect to the plus sign in the operator evolution in the Heisenberg picture. ``Time flows backwards" between the two pictures. 

This leads to the calculation in this paper that shows that the Schr\"odinger picture is as Einstein local as the Heisenberg one. We now discuss an entangling protocol between two qubits which start in the state $|+\rangle|+\rangle$. For simplicity, we will assume that the evolution is a phase kick that occurs only if the state of the qubits is $|1\rangle|1\rangle$. If again the phase is assumed to be $\pi$, the overall unitary simply transforms  $|1\rangle|1\rangle$ into $-|1\rangle|1\rangle$ and leaves all the other three basis states unchanged. 

Clearly in the Schr\"odinger picture, the final state is the maximally entangled state $|0\rangle|+\rangle + |1\rangle |-\rangle$. Up to the local Hadamard transformation of the second qubit, this state is the same as the Bell state considered before. It therefore looks Einstein non-local in the sense that if a phase is introduced now to only one of the two qubits, from the form of the state we cannot tell which qubit it is that suffered the phase. The reduced density matrices computed from the state $|0\rangle|+\rangle + e^{i\phi}|1\rangle |-\rangle$ are both again maximally mixed and contain no information about where the phase was located. 

To remedy this, we switch to the density matrix notation. Initially the density matrix is given by 
\begin{equation}
\rho = \frac{1}{2}(I+X_1)\times \frac{1}{2}(I+X_2) \; ,
\end{equation}
where, as before, $X_1= X\otimes I$ is the $X$ operator pertaining to the first qubit, while $X_2= I\otimes X$ is the $X$ operator pertaining to the second qubit. The product between the two $\times$ is just an ordinary product (not the tensorial one) so that $X_1\times X_2 = X\otimes X$ when we choose the representation above. The evolution is easily calculated since $X_1 (t)= X_1Z_2$ and $X_2(t) = Z_1X_2$.  The final state of the density matrix is therefore
\begin{equation}
\rho (t) = \frac{1}{2}(I+X_1 (t) Z_2)\times \frac{1}{2}(I+Z_1X_2 (t))\; ,
\end{equation}
which is still a product state in this new notation. The time $t$ simply indicates the state after the phase gate has been applied. 

More to the point, we see that the state of the first qubit has now acquired the dependence on the initial $Z$ component of the second qubit and the state of the second qubit has, likewise (since the unitary is symmetric with respect to the qubits), acquired the dependence of the initial $Z$ of the first qubit. It is through this mutual dependence that we see that the qubits have become entangled. They have exchanged information through the phase gate. Indeed, it is perfectly appropriate to say that each qubit measures the other one. 

Note that this does not mean that all local operations will be detectable locally. Quite the contrary. For any given initial state we have many operations that will lead to the same final state of affairs as far as local observable are concerned. This feature is intrinsic and cannot be changed by changing the notation - and it is true both in the Schrödinger and the Heisenberg picture. However the product notation we have just introduced for the Schrödinger picture allows us formally to keep track of the dynamics, just like in the Heisenberg picture. Thus we can also keep track of where local gates have been applied, by checking what factor in the product has been affected. Once more, this simply means that, in both pictures, states alone do not contain all the relevant information and we need to keep track of the dynamics (which in the Heisenberg picture is done by default by transforming the whole algebra of the relevant operators). 

Now, if we multiplied out the above state we would obtain (in the tensor product notation):
\begin{equation}
\rho (t) = \frac{1}{4}(I+X \otimes Z + Z\otimes X + Y\otimes Y) \; ,
\end{equation}
which is clearly the maximally entangled state $|0\rangle|+\rangle + |1\rangle|-\rangle$. It is this form that tricks us into believing that something non-local is going on in quantum physics. Performing the phase operation on the first qubit here would lead to the state 
$\rho (t) = \frac{1}{4}(I-X \otimes Z + Z\otimes X - Y\otimes Y)$ and this form simply does not tells us which of the two qubits was affected (since we could have obtained the same state by a suitable phase kick on the second qubit). In the product notation, on the other hand, the state would become
\begin{equation}
\rho (t) = \frac{1}{2}(I-X_1 (t) Z_2)\times \frac{1}{2}(I+Z_1X_2 (t))
\end{equation}
which exhibits the minus sign in the state pertaining to the first qubit. So if we want a fuller accounting of what is going on, the product notation is possibly therefore better than the tensor product. Even here, of course, there are operations on the second qubit that would lead to the same state, since, ultimately, the results and the information contained in the two pictures are the same. 

\section{Discussion and Conclusions}

Two points are worth summarising. One is that the state in the product notation is as local in the Schr\"odinger as it is the Heisenberg picture, meaning that the evolution of the whole can be specified by the “local factors” as we defined them. This is due ultimately to unitarity.  As we said, this does allow for different unitary dynamics to lead to the same quantum state (the same statistics), like when the phase gate is applied to one or to the other qubit in a maximally entangled state. The local and global statistics of the final state alone simply won’t tell us which of the unitaries was applied. A separate point is therefore that, for completeness, one needs to keep track of all operations. The key insight is that when we are comparing Schr\"odinger's with Heisenberg's quantum mechanics, we must compare them on an equal footing. For such comparison, it is simply wrong only to consider one state in the Schr\"odinger picture, while keeping the information about the full algebra of observables in the Heisenberg case (the latter clearly contains much more information). Of course, we could track the full basis of states in the Schrödinger picture, but this is not in the spirit of this version of quantum mechanics, as the initial state is assumed to be given as a “boundary condition”.

Two formulations of locality are relevant for this paper, the Einstein and the Bell locality. It's worth pointing out that Einstein almost certainly expected quantum theory to obey Bell locality - that seems to be the whole point of the EPR argument. Classically, there is no difference between the two, since in classical physics all quantities are represented by real numbers only, pertaining locally to the system (this is taken to be so transparent that it is not even worth discussing). However, as things stand, quantum physics satisfies the Einstein locality, but violates the Bell locality. Namely, if an action takes place in one subsystem in quantum physics, this cannot instantaneously affect any property of other subsystems (the state or the observables). Here I have argued that this is true in all representations of quantum physics and, in particular, in the Schr\"odinger picture, which is frequently (and wrongly, as shown here) seen as ``non-local".

We have been emphasising the fact that the Schr\"odinger wavefunction is, by itself, not enough. However, even here we could sidestep this problem by assuming that the given wavefunction is not just that of the system under consideration, but includes all other relevant systems. In that sense, all the actions on the system itself will be recorded in the states of the other systems participating in them. So, in our example of the maximally entangled state of two qubits undergoing a local phase kick, the knowledge about which of the qubits was kicked, would reside somewhere in the wavefunction, but not of the two qubits. Instead it could, for instance, be recorded in the quantum state of the memory of the experimentalist who implemented the transformation. This picture, akin to what is known as the Page-Wootters formulation of dynamics without dynamics\cite{Wootters,Marletto}, assumes that the wavefunction also contains all the relevant information regarding the dynamics of observables. 

The closest classical physics comes to the Schr\"odinger picture is when using the phase space representation of states and dynamics \cite{Sudarshan}. Imagine two particles elastically colliding in one-dimension but each starting with some distribution of positions and momenta $f(p_1,q_1,p_2,q_2,t_0) = f(p_1,q_1,t_0)f(p_2,q_2,t_0)$. There are no initial correlations between particles. From energy and momentum conservation it follow that 
\begin{eqnarray}
p_1 (t) & = & g_1 (p_1 (t_0),p_2 (t_0))\;\;\; p_2 (t)  =  g_2 (p_1 (t_0),p_2 (t_0)) \nonumber \\
q_1 (t) & = & q_1 (t_0) + \frac{p_1 (t)}{m} t\;\;\; q_2 (t)  =  q_2 (t_0) + \frac{p_2 (t)}{m} t\nonumber
\end{eqnarray}  
where the exact forms of the functions $g_1$ and $g_2$ are not relevant here (they can easily be calculated) and $p_1 (t), p_2 (t), q_1 (t), q_2 (t)$ are the momenta and positions of the two particles after the collision has taken place. We can see how each particle has now obtained the information about the initial momentum of the other particle. Their positions thus also become correlated and so do their densities in phase space.  The only difference with quantum mechanics is that all these quantities become q numbers, resulting in the equal time commutation relations of the form $[q_i(t),p_i(t)] = i\hbar$. The q-numbers for different particles always commute with one another. The quantum correlations arising from this could be entanglement, something that does not have any classical analogue. But this in itself is not relevant as far as Einstein's locality is concerned. The classical and quantum accounts are here identical so long as the c-numbers are replaced with the q-numbers \cite{Sudarshan}.  

There are other notions of locality in quantum physics, such as the frequently emphasised micro-causality in quantum field theory. Field operators at different space-like separated points must commute with one another (for fermions it is the quadratic Hamiltonians that commute with quadratic observables). This is a sufficient (but not a necessary) condition to ensure that no causes can propagate faster than the speed of light \cite{Eberhard}. The Einstein locality condition we used here is more general. It does not refer to any specific speed of propagation or any particular relationship between space and time. Micro-causality is therefore a stronger condition and it implies Einstein's locality. The former not only prohibits an instantaneous action at a distance, but also puts a specific speed limit (the speed of light) on any propagation. 

There is one last ``non-locality" that is frequently noted in quantum physics that deserves a mention. It is the Aharanov-Bohm effect \cite{AB}. Here the argument has frequently been made that the magnetic field acts non-locally on an interfering electron to generate a phase that affects the electron interference at a distance (i.e. where the field is zero). But, as we have shown \cite{MV}, in the fully quantum mechanical analysis, the Aharanov-Bohm phase is also acquired locally. All transformations in quantum mechanics, just as described in this paper, always take place locally. The specific formalism we use may or may not directly reflect this fact, but that is completely irrelevant since all accounts lead ultimately to one and the same set of outcomes.

The interesting question for a physicist therefore is whether there are instances where locality is truly violated and not just in some accounting procedures executed on a piece of paper while calculating. Can an action at a distance really take place instantaneously under some circumstances? Non-local extensions of quantum field theory have been considered, motivated frequently by trying to avoid various infinities, but - to the best of my knowledge - there is no evidence whatsoever that any of them are more successful than the Standard Model. More importantly, perhaps, it is questionable if one could handle such notions as ``information" or ``observables" in a theory that is Einstein non-local. A theory that violates the Einstein locality may therefore not even be testable!

\textit{Acknowledgments}: The author is grateful to Chiara Marletto and Charles B\'edard for extensive discussions and comments on issues related to this work. V.V.'s research is supported by the Moore Foundation and the Templeton Foundation.

\end{document}